\documentclass[apj]{emulateapj}

\usepackage{natbib}
\usepackage{amsmath}
\usepackage{amstext}
\usepackage{graphicx}

\usepackage{verbatim}

\newcommand{\pder}[2]{\frac{\partial#1}{\partial#2}}

\newcommand{\velvar}{v}

\usepackage[normalem]{ulem}
\newcommand{\REVaAdd}[1]{#1}
\newcommand{\REVaDel}[1]{}

\begin{document} 

\title{From Sun to interplanetary space: What is the pathlength of
  Solar Energetic Particles?}

\shorttitle{Pathlength of Solar Energetic Particles}

\author{T. Laitinen}
\affil{Jeremiah Horrocks Institute, University of Central Lancashire,
  Preston, UK}
\author{S. Dalla}
\affil{Jeremiah Horrocks Institute, University of Central Lancashire,
  Preston, UK}

\email{tlmlaitinen@uclan.ac.uk}

\shortauthors{Laitinen \& Dalla}

\begin{abstract}
  Solar energetic particles (SEPs), accelerated during solar
  eruptions, propagate in turbulent solar wind before being observed
  with in situ instruments. In order to interpret their origin through
  comparison with remote-sensing observations of the solar eruption,
  we thus must deconvolve the transport effects due to the turbulent
  magnetic fields from the SEP observations. Recent research suggests
  that the SEP propagation is guided by the turbulent meandering of
  the magnetic fieldlines across the mean magnetic field. However,
  the lengthening of the distance the SEPs travel, due to the
  fieldline meandering, has so far not been included in SEP event
  analysis. This omission can cause significant errors in estimation
  of the release times of SEPs at the Sun. We investigate the distance
  travelled by the SEPs by considering them to propagate along
  fieldlines that meander around closed magnetic islands that are
  inherent in turbulent plasma. We introduce a fieldline randow walk
  model which takes into account the physical scales associated to the
  magnetic islands. Our method remedies the problem of the diffusion
  equation resulting in unrealistically short pathlengths, and the
  fractal dependence of the pathlength of random walk on the length of
  the random-walk step. We find that the pathlength from the Sun to
  1~au can be below the nominal Parker spiral length for SEP events
  taking place at solar longitudes 45E to 60W, whereas the western and
  behind-the-limb particles can experience pathlengths longer than
  2~au due to fieldline meandering.
\end{abstract}



\section{Introduction}\label{sec:introduction}

A central goal of modelling Solar Energetic Particle (SEP) propagation
in the heliosphere is to uncover the relative timing between the SEP
production at the Sun and the remote-sensed multi-wavelength
observations of the solar eruption responsible for the SEP event. As
the interplanetary medium is permeated with a magnetic field which on
average has a Parker spiral shape \citep{Parker1958}, overlaid with
turbulence, the charged SEP transport is stochastic in nature.  The
physics of the turbulence evolution, the SEP transport parameters, and
indeed the behaviour of the charged particles in such turbulent
magnetic fields are not fully understood.

As the SEP transport is controlled by stochastic processes, it is
often modelled by use of a diffusion description, typically via a
Parker or Fokker-Planck transport equation with diffusion terms to
describe the stochasticity of the propagation
\citep[e.g.][]{Parker1965, Jokipii1966}. Several researchers have used
1D propagation models with pitch angle diffusion to to fit SEP data in
order to deconvolve the interplanetary transport from SEP observations
\citep[see, e.g.,][]{Kallenrode1993, Torsti1996, Laitinen2000,
  Droge2003, AgEa09,Gomez-Herrero2015}.

Full deconvolution of the interplanetary transport from SEP
observations, however, is complicated and usually performed only in
case studies. As an alternative, timing analysis of SEPs is often used
to connect the SEPs to the solar remote sensing observations,
particularly in large statistical studies.

In particular the velocity dispersion analysis (VDA) method is often
used to obtain the time of SEP injection near the Sun from the SEP
onset times at 1~AU \citep[e.g.][]{Lin1981, Reames1985, Torsti1998,
  KrLi00, Tylka2003, DallaEa2003Annales,
  Reames2009,Vainio2013_sepserver,Paassilta2018, Zhao2019}. In VDA, the first
particles are assumed to be injected simultaneously, and propagating
to the observing spacecraft without scattering. Under these
assumptions, the injection time at the Sun, $t_{sun}$ can be obtained from
the observed onsets by fitting
\begin{equation}\label{eq:vda}
  t_{o,j}=t_{sun}+s/v_j.
\end{equation}
to the 1-au onset times $t_{o,j}$ of the SEPs propagating with
velocities $v_j$. The pathlength $s$ in Equation~(\ref{eq:vda}) is
often expected to be the local Parker spiral length, around
1.1-1.2~au, however the statistical studies often show a very large
range of pathlengths, from $<1$~au to over 5~au
\citep[e.g.][]{Paassilta2017}. Several modelling studies have
addressed the reliability of the VDA method
\citep[e.g.][]{KaEa90,Lint04,Saiz2005,Laitinen2015onset, Wang2015},
showing that the apparent long or short pathlength may be due to the
interplanetary scattering conditions and the pre-event background,
rather than an indication of the length of the actual travelled
pathlength or energy-dependent injection time of SEPs at the Sun,
$s_{sun}(v_j)$.

Recent observations of SEP events simultaneously with multiple
spacecraft offer a different interpretation to long pathlengths.  SEP
events analysed by several authors \citep[e.g.][]{Dresing2012,
  Wiedenbeck2013, Droge2014, Cohen2014, Richardson2014} have
demonstrated the ability of SEPs accelerated near the Sun to reach a
wide range of heliographic longitudes rapidly. It has been suggested
that such a fast spread can be attributed to interplanetary
propagation of SEPs across the Parker spiral direction, modelled as
cross-field diffusion in several studies  \citep[e.g.,][]{Zhang2009,
  Droge2010, He2011, Dresing2012}. The cross-field diffusion is
believed to be dominated by random walk of the magnetic field lines,
due to turbulent fluctuations, and several theoretical approaches have
used this concept to derive spatial cross-field diffusion coefficients
\citep[e.g.][]{Jokipii1966, Matthaeus2003, Shalchi2010a,
  RuffoloEa2012}. Recent research points out that the perpendicular
propagation of the particles with respect to the mean field at short
timescales is not actually diffusive, but systematic propagation along
the stochastically meandering fieldlines.  These studies
\citep{LaEa2013b,LaEa2016parkermeand} propose to model the early SEP
propagation initially along diffusively meandering fieldlines instead,
employing a field-line diffusion approach \citep{Matthaeus1995}.

The effect of the diffusive perpendicular transport of SEPs on the
path length was investigated recently by \citet{Wang2015}, using the
3D focused transport equation \citep[e.g.][]{Zhang2009}. They found
that particles diffusing across the mean Parker spiral to wide
heliolongitudinal separations in general have longer pathlengths than
those arriving to well-connected locations at 1~au. Thus, the
cross-field propagation of particles gives a possible explanation for
the observed long pathlengts of SEPs, as given by the VDA method.

However, as we demonstrate in study, the approach using spatial
diffusion for SEP cross-field propagation has the disadvantage that it
may result in unphysically short propagation times and
pathlengths. This was recently noted by \citet{Strauss2015}, who
analysed SEP transport in the inner heliosphere using a 2D transport
equation. They found that in some cases the simulated intensities at
1~au began to rise before an unscattered SEP could have reached that
distance, that is, $t_{o,j}-t_{sun}<1\text{ au}/v_j$,
breaking causality.  As discussed by \citet{Strauss2015}, and in more
detail in our study, this is due to the fact that in diffusion
description the effect of diffusive cross-field propagation on
propagation time of the particles is not taken into account.

In this paper, we address the problem of determining the pathlength of
SEPs in \REVaDel{turbulently}\REVaAdd{the} heliospheric magnetic
field\REVaDel{. We} \REVaAdd{by analysing the length of turbulently
  meandering magnetic fieldlines, and} propose a new method for
calculating the pathlength when analysing SEP events\REVaDel{, and
  within SEP simulations, based on the physical characteristics of the
  stochastic random-walk of magnetic fieldlines that guide the SEPs in
  turbulent magnetic fields.}\REVaAdd{. Our approach is based on the
  non-linear formulation of fieldline diffusion, where the diffusion
  coefficient is proportional to the ultrascale $\tilde\lambda$
  \citep{Matthaeus1995} (instead of the correlation scale as in the
  earlier quasilinear approach \citep[e.g.][]{JokParker1968}). The
  ultrascale is identified as the size scale of turbulent magnetic
  islands \citep[e.g.][]{Matthaeus1999}, thus $\tilde\lambda$ provides
  an ideal scale for derivation of the length of turbulently
  meandering fieldlines, which control the particle propagation in
  magnetic turbulence.}

\REVaAdd{Our approach in the present study only accounts for the
  effect of particles propagating on meandering fieldlines on the
  pathlength of the particles. For consistent analysis of SEP
  propagation, our results must be implemented in an SEP transport
  model that contains parallel scattering and drifting of SEPs from
  their fieldlines due to both stochastic and large-scale gradients
  and curvatures such as those cited above. Such a model can provide
  realistic estimates for SEP events observed in the interplanetary
  space for a wide range of source and transport conditions.}

\REVaAdd{The paper is organised as follows:} In
Sections~\ref{sec:pathl-diff-equat} and~\ref{sec:pathl-stoch-diff} we
discuss the difficulties in determining the pathlength of a
stochastically propagating particle. We introduce a novel method to
determine the pathlength in Sections~\ref{sec:pathl-gauss-rand}
and~\ref{sec:stat-eval-length}, based on the scalesize of turbulent
magnetic islands that guide the random-walk of the meandering
fieldlines. In Section~\ref{sec:pathl-park-spir} we outline
simulations of stochastically meandering fieldlines in Parker spiral
geometry, and show the resulting pathlengths in
Section~\ref{sec:results}. We discuss the implications of our work in
Section~\ref{sec:discussion} and draw conclusions in
Section~\ref{sec:conclusions}.

\section{Pathlength in cartesian geometry}\label{sec:models}

\subsection{Pathlength and diffusion equation}\label{sec:pathl-diff-equat}

\REVaAdd{Propagation}\REVaDel{Cross-field diffusion} of SEPs along and across the mean magnetic field
is typically modelled using a spatial convection-diffusion
description, such as the Fokker-Planck descriptions based on works by
several authors \citep[e.g.][]{Parker1965, Zhang2009}.  Here we will
first concentrate on a very simple form of such an equation, given for
propagation of particles along constant magnetic field,
$\mathbf{B}=B_0\hat{\mathbf{z}}$, with constant velocity $\velvar$,
and diffusion across it in $x$-direction with a constant diffusion
coefficient $\kappa$. Under these conditions, the convection-diffusion
equation for the particle density $n(x, z, t)$ can be written in a
cartesian 2D form as
\begin{equation}\label{eq:diffconv}
  \pder{n(x, z, t)}{t}+\velvar\pder{n(x,z,t)}{z}=\kappa\pder{^2n(x,z,t)}{x^2},
\end{equation}
In this simple model, pitch angle diffusion, which would
give rise to diffusion of the particles along the magnetic field
direction, is ignored.

Equation~(\ref{eq:diffconv}) and its analytical solution,
\begin{equation}\label{eq:twodsol}
  n(x, z, t)= \frac{n_0(z-\velvar t)}{2 \sqrt{\pi \kappa t}}\mathrm{e}^{-x^2/(4\kappa t)},
\end{equation}
for an impulsive point-injection $n_0=\delta(t)\delta(x)\delta(z)$,
represent an asymptotically valid solution for a random-walk process
across the field with a large number of steps, $N\gg 1$, of particle
population that propagates along the field with veloctity $\velvar$, and
diffuses across the field in $x$-direction. However, this solution is
unphysical in that the diffusion across the field is not limited by
the particle velocity: at a given $z=\velvar t$ the density is non-zero at
all $x$-values.

This unphysical nature of the solution can perhaps be better
demonstrated when viewing the solution with the stochastic
differential equation (SDE) approach. Diffusion and
diffusion-convection equations can be solved using SDEs
\citep[e.g.][]{Zhang1999, Gardiner2009, Strauss-Effenberger-2017}, by
use of statistics derived from pseudoparticles that are propagated as
given by SDEs that are equivalent to the diffusion equation. In the
case of our simple model given by Equation~(\ref{eq:diffconv}), the
corresponding SDE equations are
\begin{equation}\label{eq:SDEeqs}
\begin{aligned}
  dx &= \sqrt{2\kappa dt}W\\
  dz &= \velvar dt.
\end{aligned}
\end{equation}
where $W$ is a Wiener process, described as a Gaussian random number
with unity variance and zero mean\footnote{Note that the Wiener
  process is often formally written as having $\left<W(t)^2\right>=t$,
  or $\left<dW^2\right>=dt$. However, in applications it is often more
  convenient to consider the dependence of the process on time
  separately.}. The solution given by Equations~(\ref{eq:SDEeqs}) is
equivalent to Equation~(\ref{eq:twodsol}). We can easily see that
after a time $t$ the pseudoparticles solved with
Equation~(\ref{eq:SDEeqs}) are at $z=\velvar t$, and spread along the
$x$ with variance $\left<x^2\right>=2\kappa t$. However, physically
the particle propagating with velocity $\velvar$ can only have
propagated a maximum distance of $s=\velvar t$ in time $t$. Thus, the
distance the particles have propagated along the x-axis is not taken
physically into account in Equation~(\ref{eq:diffconv}).

To develop discussion into how the unphysicality of the diffusion
equation can be taken into account we define the pathlength $s$ of a
particle integral of 
\begin{equation}
  \label{eq:pathlendef}
  ds = v dt,
\end{equation}
that is, the pathlength of the particle is defined as the distance a
particle propagates with velocity $v$ in time $dt$. As can be seen in
Equation~(\ref{eq:SDEeqs}), for the \emph{diffusion solution} the
pathlength of the particle is given as
\begin{equation}
  s_{diff} = \int_P dz,
\end{equation}
What this means is that as the particle propagates along a stochastic
path $P$, the distance the particle diffuses across the
field, $dx$, does not ``consume time'', and thus according to the
definition of Equation~(\ref{eq:pathlendef}), does not contribute to
the pathlength. The practical consequence of this is that if we
consider the arrival time of a particle from, say, the origin to a
point $(X, Z)$, the solution of Equation~(\ref{eq:diffconv}) gives
$t=\velvar Z$, thus a too-early arrival time compared even to a
non-diffusing particle propagating along direct path, for which
$t=\velvar \sqrt{X^2+Z^2}$.

\subsection{Pathlength and stochastic differential equations}\label{sec:pathl-stoch-diff}

As discussed above, the solution of the diffusion-convection equation,
Equation (\ref{eq:diffconv}), gives too-early onsets, or too short
pathlengths, for particles with a finite velocity. However, the SDE approach
to solving the diffusion-convection equation provides an opportunity to
estimate the distance propagated by the particle as an \emph{SDE steplength}
\begin{equation}
\delta s_{SDE} = \sqrt{\delta x^2 + \delta z^2}, \label{eq:sdelenmethod}
\end{equation}
that is, calculating the length of each stochastic 2D step that can
then be used to evaluate the time required for taking the step,
$\delta t=\delta s/v$.

Let us investigate this approach further. Using
Equations~(\ref{eq:SDEeqs}) we obtain
\begin{equation}
  \delta s_{SDE} = \velvar\delta t \left(\frac{2\kappa W^2}{\velvar^2\delta t}+1\right)^{1/2}.\label{eq:SDElen}
\end{equation}
We can define $T_z\equiv N\delta z/\velvar \equiv Z/\velvar$, where $Z$ is the
distance propagated along $z$-direction in time $T_z$, and $T_z=N\delta t$
is the corresponding time, excluding any contribution from stepping in
$x$-direction. Using these, we have 
\begin{equation}
  \label{eq:SDElenN}
  \delta s_{SDE} = \velvar\frac{T_z}{N}\left(\frac{2\kappa W^2}{\velvar^2 T_z}N+1\right)^{1/2}.
\end{equation}
We can now evaluate the pathlength of the particle due to $N$ SDE steps,
estimating $W^2$ as unity\footnote{Note that the
  Equation~(\ref{eq:SDElen}) could be further developed using It\^o
  calculus \citep[e.g.][]{Gardiner2009}. However, it is easy to see
  that the integral diverges at the limit of $\delta t\rightarrow 0$,
  rendering use of It\^o calculus not useful.}, to be
\begin{equation}
  \label{eq:SDEtotlen}
  s_{SDE}\sim \velvar T_z\left(\frac{2\kappa}{v^2 T_z}N+1\right)^{1/2}.
\end{equation}
Thus, the time required for the particle to propagate the path would
be 
\begin{equation}\label{eq:SDEtottime}
  T_{SDE}=\frac{s_{SDE}}{\velvar}\sim T_z\left(\frac{2\kappa}{\velvar^2 \delta t}+1\right)^{1/2}.
\end{equation}

As can be clearly seen in Equations~(\ref{eq:SDEtotlen})
and~(\ref{eq:SDEtottime}), the approach for estimating the pathlength
using Equation~(\ref{eq:sdelenmethod}) results in an unphysical
solution, as the obtained distance $s_{SDE}$, and consequently the
propagation time $T_{SDE}$, depends on the selected timestep length,
$\delta t$. For large timesteps, the total propagation time given by
Equation~(\ref{eq:SDEtottime}) approaches $T_z$, that is, it has no
contribution from the diffusive steps taken across the $z$
direction. At small timesteps, the propagation time scales as
$\delta t^{-1/2}$, or $N^{1/2}$, approaching infinity. The propagation
time scaling as $N^{1/2}$ is consistent with the fractal dimension
$D=2$ of the path of a particle in Brownian motion
\citep[e.g.][]{Mandelbrot1982, Rapaport1985}.

\begin{figure}
  \centering
  \includegraphics{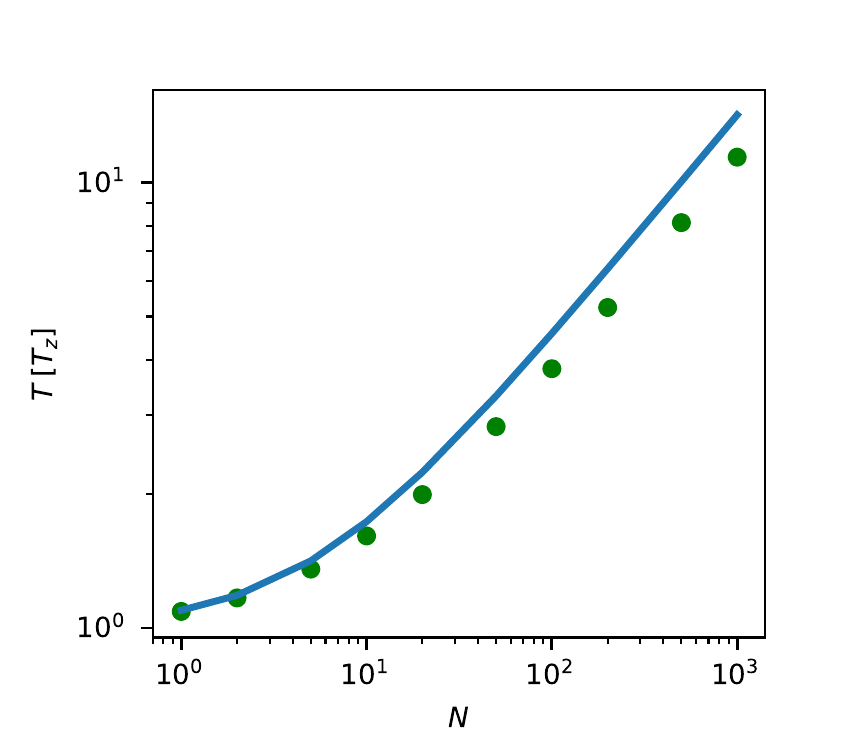}
  \caption{Dependence of propagation time $T_{SDE}$ on the number of
    steps $N$, for $\kappa=0.1$ for total time
    $T_z\equiv N \delta t=1$. The curve shows the length given by
    Equation~(\ref{eq:SDEtottime}), and the symbols show the
    results of SDE simulations, where at each step the distance
    propagated is calculated from
    Equation~(\ref{eq:sdelenmethod}).\label{fig:cart_Nvspathlen}}
\end{figure}

We demonstrate the dependence of the pathlength of a random-walking
particle on the steplength further by simulating pseudo-particles
using the SDE Equations~(\ref{eq:SDEeqs}) and calculating the
pathlength with Equation~(\ref{eq:sdelenmethod}). In
Figure~\ref{fig:cart_Nvspathlen}, we show the resulting $T=s/\velvar$ for
parameter values $\kappa=0.1$, $\velvar=1$ and $T_z=1$. with the filled
circles showing the mean $T$ for 10,000 pseudoparticles as a function
of the number of timesteps, $N$. The solid curve shows the analytical
expression given by Equation~(\ref{eq:SDEtottime}). The simulations
show clearly both the asymptotic $N^{1/2}$ at large $N$, and the
approach to unity at small $N$, as predicted by
Equation~(\ref{eq:SDEtottime}).

Thus, it appears that the method of using
Equation~(\ref{eq:sdelenmethod}) for evaluating the pathlength of the
particles provides an unphysical result: The pathlength depends on how
we select the timestep lengths. In
general, we are not free to determine this timestep arbitrarily. It is
typically determined so that the number of timesteps is large, to
ensure $N\gg 1$ and to obtain sufficiently large statistical
distribution of the steps. The timestep is also limited by the
possible spatial and temporal variation of the diffusion coefficient
(and other terms such as the background magnetic field in heliospheric
magnetic field configuration): such properties should not change
appreciably during the SDE step.

It is easy to see that the problems arising from using
Equation~(\ref{eq:sdelenmethod}) for pathlength determination are not
limited to our simple diffusion-convection
Equation~(\ref{eq:diffconv}): Similar results can be derived also for
2- or 3-dimensional problems where propagation in one or more
directions is diffusive. For spatial diffusion in two cartesian
directions $x$ and $z$, the pathlength would be
\begin{equation}
  s_{SDE,2D}\sim T_d \left(\frac{2\kappa_x+2\kappa_z}{ \delta t}\right)^{1/2}
\end{equation}
where $T_d=N\delta t$. \REVaAdd{Thus, the pathlength depends on
  $\delta t$ also for 2- and 3-dimensional spatial diffusion in SDE
  picture. This is a direct consequence of the pathlength of a
  random-walking particle being fractal, which results in infinite
  pathlength for an infinitesimal step size
  \citep[e.g.][]{Mandelbrot1982}.}

\REVaAdd{However, as noted by \cite{Rapaport1985}, in real physical
  situations the path-length of a random-walking particle is not
  infinite, but limited by the physics behind the random-walking
  process. Thus, in order to evaluate the pathlength of particle
  propagating in stochastic magnetic fields, we must understand the
  physics behind the random walk.}

\subsection{Pathlength and Gaussian random walk of magnetic field
  lines: \REVaAdd{Turbulent islands and ultrascale}}
\label{sec:pathl-gauss-rand}

The evaluation of the pathlength from the SDE steps in $x$ and $z$
directions, as given by Equation~(\ref{eq:sdelenmethod}), proved
unphysical. However, it does provide a possibility to
solve the problem of determining the pathlength of a diffusing
particle, given a suitable physical framework. Here, we employ
field-line random walk as the framework for determining the pathlength of
a diffusing particle.

In the SDE method, the stepsize does not have physical meaning, but in
the physical world it does. Particle cross-field diffusion is believed
to be dominated by their following the turbulent random-walk of the
magnetic field lines \citep[e.g.][]{Fraschetti2011perptimetheory}. The
fieldlines do not, however, meander at infinitesimal scales, since
physical processes damp small-scale fluctuations. Thus, a length scale
that would describe the meandering of the field lines is a good
candidate for evaluation of the pathlength of a particle in turbulent
magnetic fields.

Such a length scale can be derived from the definition of a field
line diffusion coefficient and the concept of an ultrascale,
$\tilde\lambda$. The field line diffusion coefficient for 2D-dominated
turbulence is given by
\begin{equation}
  \label{eq:dfl}
  D_{FL}=\frac{\tilde\lambda\sqrt{\delta B_\perp^2/2}}{B}
\end{equation}
where $B$ is the magnitude of the ambient background magnetic field,
and $\delta B_\perp^2$ is the turbulence variance
\citep{Matthaeus1995}. For 2D turbulence spectrum $S(k)$, the
ultrascale is defined as
\begin{equation}
  \tilde\lambda^2=\frac{\int S(k)k^{-2}\mathrm{d}k}{\delta B^2}.
\end{equation}
where $k$ is the wavenumber.

\citet{Matthaeus1999} gives the ultrascale an interpretation as the
representative scale size of turbulent closed magnetic 2D structures,
``magnetic islands'', in the cross-field direction, $x$. The fieldlines
in 2D-dominated turbulence can be thought to be either trapped in
magnetic islands or meandering freely around these islands
\citep{Ruffolo2003, Chuychai2007}. We can thus consider the ultrascale
to be the relevant cross-field length scale for the meandering of the
untrapped fieldlines around the islands that are of size
$\tilde\lambda$.

The distance $\Delta z_{FL}$ along the mean field direction as the field line
propagates a cross-field length $\tilde\lambda$ can then be evaluated
using the fieldline diffusion coefficient,
Equation~(\ref{eq:dfl}), using general definition of  diffusion
coefficient $D_{FL}=\left<\Delta x_{FL}^2\right>/(2\Delta z_{FL})$, where the
distance along $z$-axis takes place of time in the denominator for
field line diffusion. If we consider the mean square cross-field step
given as the ultrascale, $\left<\Delta x_{FL}^2\right>=\tilde\lambda^2$, we
can write 
\begin{equation}\label{eq:dlfromultra}
  \Delta z = \frac{\tilde\lambda^2}{2 D_{FL}}=
  \tilde\lambda\frac{B}{\sqrt{2\delta B^2}}.
\end{equation}
Equation~(\ref{eq:dlfromultra}) gives a natural interpretation to the
fieldline diffusion coefficient in Equation~(\ref{eq:dfl}): the field
line random walk across the mean field direction is described as
random walk with step size $\tilde\lambda$, with the ratio between
the steps along and across the field, $\Delta z_{FL}/\Delta x_{FL}$, equal to
$B/\delta B_\perp$.

Using the step length as defined by the turbulent island size, given
by Equation~(\ref{eq:dlfromultra}), we can solve the pathlength of the
meandering fieldline as \emph{Gaussian random walk} with
\begin{equation}
  \label{eq:gaussianrdnwalk}
  \Delta x_{FL} = \sqrt{2 D_{FL} \Delta z_{FL}} W,
\end{equation}
The pathlength of a particle following such a
fieldline can then be estimated integrated using equation
\begin{equation}\label{eq:grwlength}
  s_{FL} = \sum_i \sqrt{\Delta z_{FL,i}^2 + \Delta x_{FL,i}^2},
\end{equation}
with the steps along and across the field given by
Equations~(\ref{eq:dlfromultra}) and~(\ref{eq:gaussianrdnwalk}),
respectively. \REVaAdd{Analogously, the pathlength of a particle
  following a fieldline meandering around turbulent magnetic islands
  without scattering has a pathlength
\begin{equation}\label{eq:grwlength}
  s = \sum_i \sqrt{\Delta z_{i}^2 + \Delta x_{i}^2},
\end{equation}
where $\Delta z_i=\Delta z_{FL,i}$ and $\Delta x_i=\Delta
x_{FL,i}$. In the following, we will drop the subscript \emph{FL} for
convenience, with the symbols prepended with $\Delta$ referring to
paths due to meandering aroung magnetic islands.}

\subsection{Statistical evaluation of the length of a meandering path}\label{sec:stat-eval-length}

To estimate the length of a meandering field line, it is useful to
derive an expression that uses statistical properties of the
turbulence giving rise to the meandering of fieldlines. Furthermore,
we are usually interested on the pathlength of the particles to a
given location in space, such as Earth, relative to the particle
source. Here we will derive an expression for pathlength as a function
of the observer coordinates relative to the particle source and
turbulence properties, for our cartesian geometry case with constant
background magnetic field.

Consider a path of a particle from the origin $(0, 0)$ to some point
$(X, Z)$, due to the Gaussian random walk process. The step across the
mean field is given by Equation~(\ref{eq:gaussianrdnwalk}). The mean
pathlength $\left<s\right>$ would then be the mean length given by
Equation~(\ref{eq:grwlength}) of all possible paths between the origin
and $(X, Z)$.

To evaluate the pathlength, we decompose the cross-field step to a
systematic part, $\Delta x_{a, i}$ which will move the particle the
cross-field distance $X=\sum_i\Delta x_{a,i}$, and a stochastic part
$\Delta x_{s,i}=\sqrt{2 D_{FL}\Delta z_i} W_i$ with $\left<\Delta x_{s,i}\right>=0$,
and $\left<\Delta x_i^2\right>=2D_{FL}\Delta z_i$. With these definitions, the
step length is given as
\begin{equation}\label{eq:sdelenxai}
  \Delta s_i=\sqrt{\Delta z_i^2+\left(\Delta x_{a,i}+\sigma_i W_i\right)^2}
\end{equation}
where $\sigma_i=2 D_{FL}\Delta z_i$. We can further define the length
of the systematic step, $(\Delta x_{a,i}, \Delta z_i)$, as
\begin{equation*}
\Delta s_{0,i}=\sqrt{\Delta z_i^2+\Delta
  x_{a,i}^2},
\end{equation*}
noting that $s_0\equiv\sum s_{0,i}=\sqrt{X^2+Z^2}$ is the distance
between $(0,0)$ and $(X,Z)$ along straight line.

Using the notations give above, we expand Equation~(\ref{eq:sdelenxai}) to
second order in $\sigma_i W_i$ to give 
\begin{equation}
  \Delta s_i\approx \Delta s_{0,i} + \frac{\sigma_i\Delta
  x_{a,i}}{ \Delta s_{0,i}} W_i+\frac{\sigma^2\Delta z_i^2}{2 \Delta s_{0,i}^3}W_i^2.
\end{equation}
Averaging this over the Wiener process $W$, and noting that
$\left<W\right>=0$ and $\left<W^2\right>=1$, we find the mean length
of the step
\begin{equation}\label{eq:lengthestimate}
    \left<\Delta s_i\right> \approx \Delta s_{0,i} \left(1 + \frac{D_{FL}\Delta z_i^3}{\Delta s_{0,i}^4}\right).
  \end{equation}
  Substituting from Equations~(\ref{eq:dfl})
  and~(\ref{eq:dlfromultra}), we get a simpler form,
  \begin{equation}
    \label{eq:lengthvsturb}
    \left<\Delta s_i\right> \approx \Delta s_{0,i} \left(1 +
      \frac{\delta B^2}{B^2}\frac{Z^4}{s_{0}^4}\right).
  \end{equation}
  If we further assume that $\delta B^2/B^2$  is constant, we find for
  the mean pathlength
  \begin{equation}
    \label{eq:fulllengthvsturb}
    \left<s\right> \approx s_0 \left(1 +
      \frac{\delta B^2}{B^2}\frac{Z^4}{s_{0}^4}\right).
  \end{equation}
  Note that the term $Z/s_0=\cos\alpha$, where $\alpha$ is the angle
  between the $z$-axis and the line connecting the origin and the
  point $(X, Z)$. Thus, the term $(Z/s_0)^4$ is~1 for $X=0$ and
  decreases to~0 for larger values of $|X|$.

  This form is beneficial in that it depends only on the statistical
  properties of the turbulence, and it doesn't depend on the variables
  describing the step length. It should be noted that this analysis is
  valid only for $\sigma<\Delta s_0$, that is, $D_{FL}<\Delta z$
  which, according to our definitions in Equations~(\ref{eq:dfl})
  and~(\ref{eq:dlfromultra}) holds for $\delta B^2< B^2$, a valid
  assumption in the inner heliosphere.

\begin{figure}
  \centering
  \includegraphics{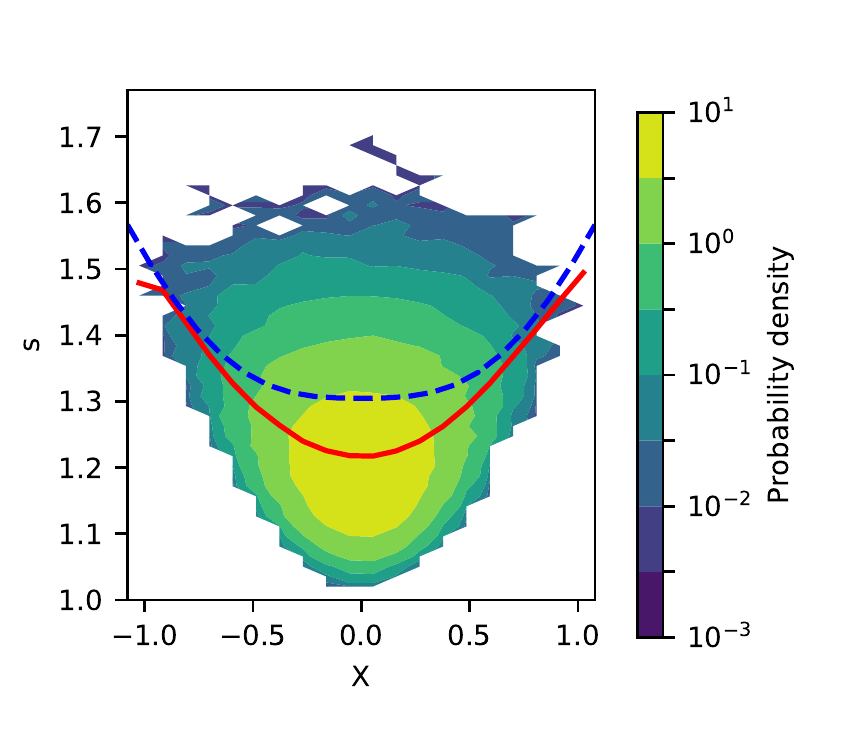}
  \caption{Dependence of path length $s$ on the total cross-field
    deviation $X$ after $N=10$ Gaussian random walk steps, with
    $D_{FL}=0.03$~au and the integrated distance along the mean field
    direction, $Z=1$~au. The contours show the probability density of
    the pathlengths obtained from Gaussian random walk simulations,
    and the solid red curve the pathlength the mean pathlength of the
    simulated particles as a function of $X$. The blue dashed curve
    shows the result of
    Equation~(\ref{eq:fulllengthvsturb}). \label{fig:cart_crossdev_vs_length}}
\end{figure}

  In Figure~\ref{fig:cart_crossdev_vs_length}, we plot the mean
  pathlength as given by Equation~(\ref{eq:fulllengthvsturb}) together
  pathlengths derived from SDE simulations of magnetic field line
  meandering, with the pathlength calculated using
  Equation~(\ref{eq:grwlength}). In the simulations, paths are started
  from the origin, and propagated until they reach distance $Z=1$~au
  along the fieldline, with the pathlength calculated as sum of
  lengths given by Equation~(\ref{eq:sdelenxai}).  In
  figure~\ref{fig:cart_crossdev_vs_length} the contours represent the
  probability \REVaDel{distribution}\REVaAdd{density} of simulated
  pathlengths as a function the final position $(X, Z=1)$, for
  simulation parameters $N=10$ and $D_{FL}=0.03$~au, corresponding to
  the values in \citet{LaEa2016parkermeand} at 1~au, resulting in
  $\sigma=0.078$~au. The dashed blue curve shows
  Equation~(\ref{eq:fulllengthvsturb}), whereas the solid red curve
  shows the mean pathlength obtained from the simulations, as a
  function of $X$. As we can see, the mean pathlength is
  well-reproduced by the estimate, thus similar estimates could be
  used to analyse the pathlength also in more complicated
  scenarios. The shortest \REVaDel{distance}\REVaAdd{distances} in
  Figure~\ref{fig:cart_crossdev_vs_length} followREVaDel{s} well the
  length of a direct path between the origin and $(X, Z)$,
  $s_0=\sqrt{X^2+Z^2}$.\REVaAdd{ The shortest pathlength for statistics of
  100,000 paths is 1.013~au.}

\section{Pathlength in Parker spiral configuration}
\label{sec:pathl-park-spir}

We will now consider the length of the meandering path in the context
of a Parker spiral field. We limit this study to 2D, in the
heliographic equatorial plane, however the same method can easily be
extended to 3D. In 2D, the parker spiral can be represented as a polar
Archimedean curve
\begin{equation}\label{eq:parkerspiral}
r=\phi_0+a\phi,
\end{equation}
where $r$ is the heliocentric distance, $\phi$
heliolongitude and $a=V_{sw}/(\Omega_0\sin\theta)$, with
$\Omega_0=-2.86533\times 10^{-6}\, \mathrm{rad/s}$ the solar rotation
rate, $\theta=90^\circ$ the colatitude at the heliographic equator, and $V_{sw}$ the solar wind speed. We use $a=-1$, which corresponds to $V_{sw}=430$~km/s.

Within the simulations presented below, the paths are traced in a
locally cartesian frame with one axis along the Parker spiral, with
stochastic steps across the Parker spiral direction. This is the
method adopted in several studies SEP transport is analysed analysed
by solving a 3D particle transport equation with SDE equations
\citep[e.g.][]{Zhang2009,Droge2010}.

We use a field line diffusion coefficient \REVaAdd{$D_{FL}$} similar
to that in \citet{LaEa2016parkermeand}. However, in this paper we use
an analytic formulation based on the ultrascale
\REVaAdd{$\tilde\lambda$ \citep{Matthaeus2007}}, given in
Appendix~\ref{sec:dfl_analytic}, \REVaAdd{instead of integrating the
  turbulence spectrum as in \citet{LaEa2016parkermeand}. Both
  approaches are consistent with \citet{Matthaeus1995}}.

As in Section~\ref{sec:models}, we will consider three methods for
calculating the pathlength.
\subsection{Pathlength and diffusion solution}

The SDE for diffusively random-walking field line is given
as\footnote{It should be noted that depending on the physics of the
  underlying processes, a term proportional to the gradient of
  $D_{FL}$ (or divergence of the diffusion tensor) is typically
  included in Equation~(\ref{eq:sde_diff_parker}). However, we neglect
  it as a small term compared to $dl$ and $dr_{l,\phi}$.}.
  \begin{equation}
    \label{eq:sde_diff_parker}
    dr_{l,\phi} = \sqrt{2 dl D_{FL}(r)} W,
  \end{equation}
  where $dl$ is a step along the local Parker spiral direction, and
  $dr_{l,\phi}$ a step normal to the Parker spiral in equatorial
  plane.  As discussed in Section~\ref{sec:pathl-diff-equat}, the
  cross-field steps do not contribute to the propagation time under
  diffusion description. Thus, for the diffusion solution case in
  Parker spiral configuration, the pathlength is given as
  \begin{equation}
    \label{eq:len_diff_parker}
    s_{diff}=\int_P dl
  \end{equation}
  where $P$ is the path determined by
  Equation~(\ref{eq:sde_diff_parker}).

  Within the numerical solution of Equation(\ref{eq:sde_diff_parker})
  the SDE step along the Parker spiral, $dl$, is limited by the
  variation of the diffusion coefficient $D_{FL}$, as well as the
  changing geometry of the system as the path meanders across the
  Parker spiral geometry: none of $D_{FL}(r)$, the Parker spiral
  direction, nor the direction across the local Parker spiral can be
  allowed to change appreciably during the step given by
  Equation~(\ref{eq:sde_diff_parker}). We have chosen to use the scale
  length of the magnetic field,
  $L_B=B/(\partial B/\partial r)\sim 2 r$, as a representative scale
  of change of the inner heliosphere, and use $\Delta l=0.01\, r$ so
  that the changes in the background medium would be small within the
  SDE step.

  To solve the field line path, we use a leapfrog scheme, where the
  magnitude and direction of the $dr_{\perp,\phi}$ step is
  evaluated at the midpoint between two consecutive steps along the
  Parker spiral.

  \subsection{Pathlength and stochastic differential equations}

  Solution of the SDE steplength, as defined in
  Section~\ref{sec:pathl-stoch-diff}, is given for Parker spiral by
  Equations
  \begin{align}
    \delta r_{l,\phi} &= \sqrt{2 \delta l D_{FL}(r)} W. \label{eq:sde_parker}\\
    s_{SDE}&=\sum_i \sqrt{\delta l_i^2+\delta r_{l,\phi,i}^2}. \label{eq:len_sde_parker}
  \end{align}
  The SDE given by Equation~(\ref{eq:sde_parker}) is equivalent to
  diffusion case, Equation~(\ref{eq:sde_diff_parker}), only the
  determination of the pathlengths differ. \REVaDel{Thus, the}\REVaAdd{The} integration
  scheme\REVaDel{, and the determination of the step size along the Parker
  spiral,} is the same as in the first case. \REVaAdd{Likewise, we use
  the same steplength as in the first case, $\delta l_i=0.01\,r$.}

  \subsection{Pathlength and Gaussian random walk of fieldlines}
  
Solution of the Gaussian random walk step length in Parker geometry is
given by
  \begin{align}
    \label{eq:parkersteplen}
    \Delta l &= \frac{\tilde\lambda^2}{2D_{FL}}\\
    \Delta r_{l,\phi} &= \sqrt{2 \Delta l D_{FL}(r)} W. \label{eq:grw_parker}\\
  s_{GRW}&=\sum_i\sqrt{\Delta l_i^2+\Delta r_{\perp,\phi}^2}.\label{eq:grw_len_parker}
\end{align}
For the ultralength, we use $\tilde\lambda=\sqrt{\lambda_c L}$ (see
Appendix~\ref{sec:dfl_analytic}), with $\lambda_c=0.007$~au and and
$L=r$, as in \citet{LaEa2016parkermeand}. As both $\tilde\lambda^2$
and $D_{FL}$ are proportional to $r$ for most of the space inside 1~au
in our model, the Equation~(\ref{eq:dlfromultra}) results in a roughly
constant meandering length scale $\Delta l=0.1$~au. We note that our
value of $\lambda_c$ results in ultrascale $\tilde\lambda=0.08$~au at
1~au, consistent with the simulation results in \citet{Ruffolo2003},
who discussed their simulations with $\tilde\lambda=0.06$~au in the
context of SEP intensity dropouts over scales $\sim 0.03$~au. Flux
ropes of similar scales have also been observed in situ in the
heliosphere, with \citet{Yu2016} finding median size of 0.02~au for
small-scale fluxropes at STEREO spacecraft.

As the step $\Delta l=0.1$~au is quite long and may cause numeric
errors, we integrate the pathlength as in the previous two cases, but then
interpolate the $(r, \phi)$ coordinates at distances
$\Delta l=0.1$~au. Other methods, such as smoothing the path with an
appropriate kernel of length determined by
Equation~(\ref{eq:dlfromultra}) before integrating the length can be
also used.

\section{Results} \label{sec:results}

We use the model presented in Section~\ref{sec:pathl-park-spir} to
study the length of meandering field lines in the heliosphere. The
paths are started from a point at the solar surface, at
($r=r_\odot, \phi=0$).

\begin{figure}
  \centering
  \includegraphics{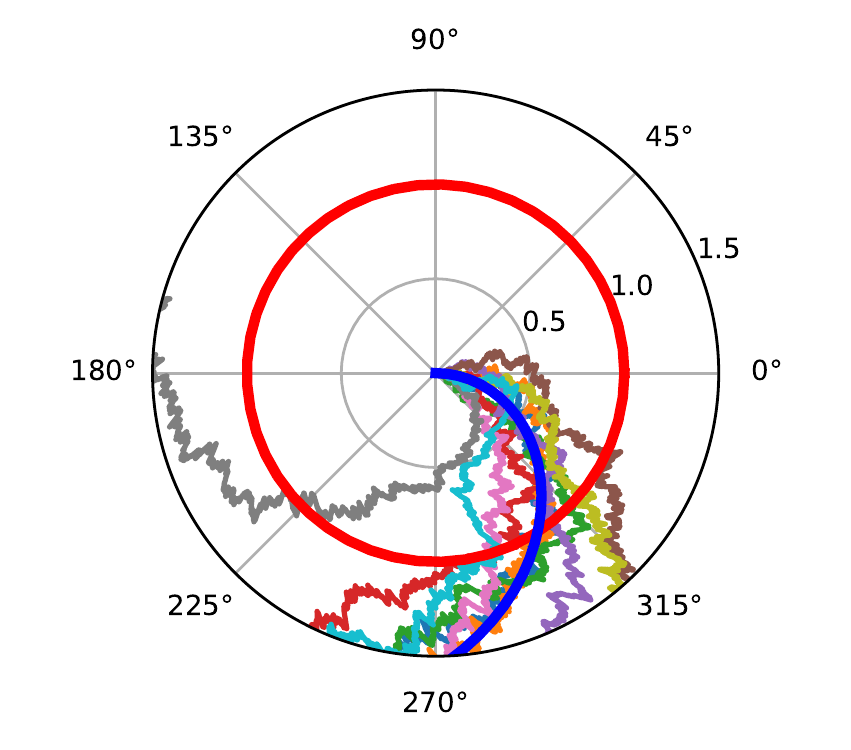}
  \caption{A sample of stochastically meandering field lines,
    simulated with step length $dl=0.01\,r$~au along the Parker spiral,
    and field-line diffusion coefficient given by
    Equation~(\ref{eq:dflparker}). The thick red curve is at 1~au
    radial distance from the Sun, and the thick blue curve shows the
    Parker spiral for solar wind velocity
    $V_{sw}=430$~km/s.\label{fig:parkerfieldcontour}}
\end{figure}

In Figure~\ref{fig:parkerfieldcontour}, we show a sample of meandering
paths in the inner heliosphere, obtained from our model. The thick
blue curve depicts the Parker spiral starting from longitude $\phi=0$,
which crosses the 1~AU distance (red circle) at longitude
$\phi=-1$~rad, or $303^\circ$. It should be noted that the meandering
paths can cross the 1~AU sphere several times, and from both inside
and outside of Earth's orbit, due to the curving of the Parker spiral.

To analyse the pathlengths, we follow the field lines to a total
distance along the Parker spiral of $l=5.6$~au. Each time the path
crosses radial distance of 1~au from the Sun, we record the
heliolongitude of the crossing and the length the meandering path as
defined in Equations~(\ref{eq:len_diff_parker}),
(\ref{eq:len_sde_parker}) and (\ref{eq:grw_len_parker}), for the three
methods.

\begin{figure*}
  \centering
  \includegraphics{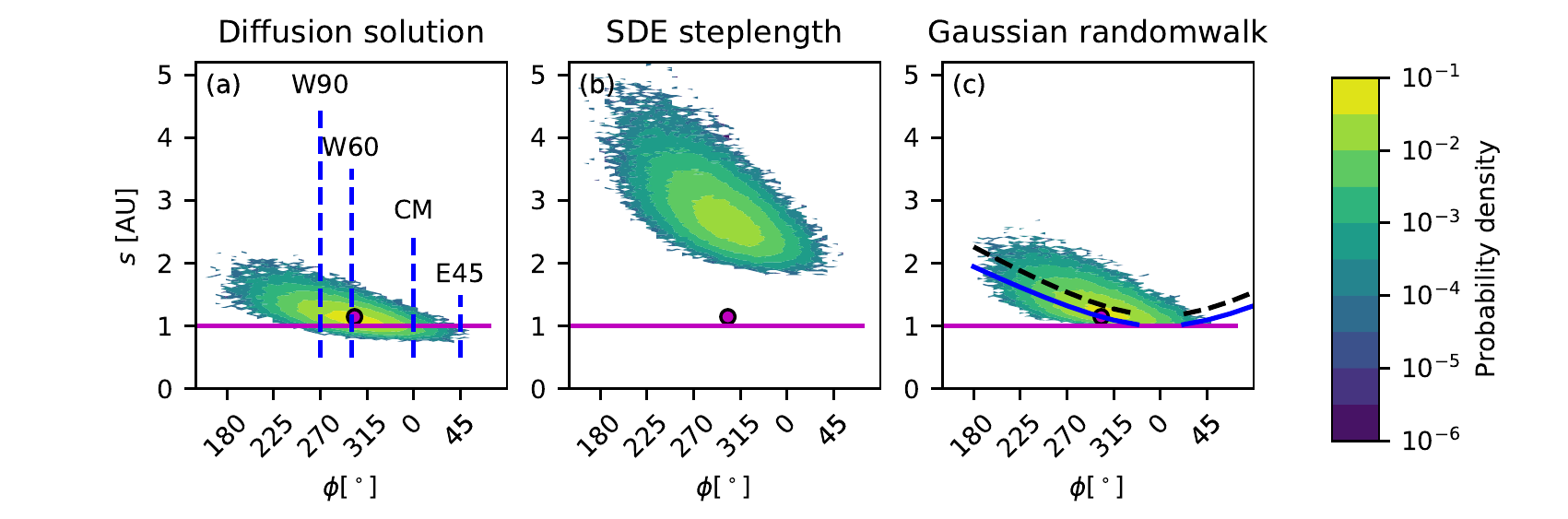}
  \caption{The probability density of pathlengths $s$ as a function of
    heliolongitude, $\phi$, at heliocentric distance of 1~au. The
    pathlengths are integrated (a) as the sum of the step lengths
    along the Parker spiral (Equation~(\ref{eq:len_diff_parker}); (b)
    sum of SDE step lengths using Equation~(\ref{eq:len_sde_parker})
    with $\delta l = 0.01 r$; and (c) with Gaussian random walk using
    Equation~(\ref{eq:grw_len_parker}), with the step length given by
    Equation~(\ref{eq:parkersteplen}). The horizontal line shows
    pathlength of 1~au, and the magenta-filled circle is at
    $\phi=303^\circ$, the longitude connected to the source longitude
    $\phi=0^\circ$ at the Sun, and $s=1.15$~au, the nominal Parker
    spiral length. The blue vertical dashed lines in panel (a)
    correspond to Solar source longitudes as viewed by an observer at
    Earth. The solid blue curve and dashed black curve in panel (c)
    show the estimated mean pathlength using
    Equation~(\ref{eq:parkerestimate}) for $\rho=0$ and $\rho=1$,
    respectively.}\label{fig:alongparker_vs_pathlen}
\end{figure*}

In Figure~\ref{fig:alongparker_vs_pathlen}, we show the propability
density of the integrated pathlength of the meandering paths at 1~au
radial distance, as a function of heliolongitude. The horizontal line
shows pathlength of 1~au, whereas the magenta-filled circle is at the
longitude and pathlength for the nominal Parker spiral connected to
$\phi_0=0^\circ$ at the Sun along the nominal Parker spiral for
$a=-1$, that is $\phi=303^\circ$ and $l=1.15$~au. The labelled blue
vertical dashed lines describe the solar longitude of the source as
would seen by an observer at Earth. The label CM, at
$\phi=0^\circ$, corresponds to a source at the centre of solar disc,
centre meridian, whereas W60 depicts a well-connected western source
and E45 a poorly-connected eastern source. The label W90 represents
the western limb, thus longitudes on its left side represent
connection to backside events.

In panel (a), we show the pathlength calculated as given by the
\emph{diffusion solution}, as given by
Equation~(\ref{eq:sde_diff_parker}), that is, just taking into account
the distance propagated along the Parker spiral, corresponding to the
SDE solution of spatial cross-field diffusion of field lines, as
discussed in Section \ref{sec:pathl-diff-equat}. As can
be clearly seen, the shortest pathlengths are shorter than the
distance from the Sun to 1~au (horizontal line). This corresponds to
the unphysically early SEP onset times in SEP transport simulations
with spatial diffusion, which was discussed in \citet{Strauss2015}.

In panel (b) of Figure~\ref{fig:alongparker_vs_pathlen}, we show the
pathlength of a meandering fieldline in the Parker spiral geometry as
calculated with \emph{SDE steplengths}, with
Equation~(\ref{eq:len_sde_parker}). While these pathlengths are not
unphysically short as in panel (a), they are very long, contradicting
observations analysed with VDA method \citep[e.g.][and other
references cited in
Section~\ref{sec:introduction}]{Paassilta2017}. As discussed in
Section~\ref{sec:pathl-stoch-diff}, this is an artificial feature due
to the fractal nature of the pathlength of random walk, which results
in unphysical dependence of the pathlength on the adopted stepsize,
\REVaAdd{$\delta l=0.01\,r$}.

We now turn to using the concept of Gaussian random walk of magnetic
field lines discussed in Section~\ref{sec:pathl-gauss-rand}, where we
derived a physically-meaningful scale length for the meandering of
fieldlines. We show the probability density of the pathlengths in
Figure~\ref{fig:alongparker_vs_pathlen}~(c), as calculated with
Equation~(\ref{eq:grw_len_parker}). As can be seen, for the Gaussian
random walk case the pathlengths range between 1 and 3~au at all
heliolongitudes, with well-connected ($\phi=303^\circ$, magenta-filled
circle) longitudes having pathlengths ranging from 1~to 2~au, with the
most probable pathlenth slightly longer than the nominal 1.15~au for
the 430~km/s solar wind. For events occuring on most parts of the
solar disk our result suggest that the pathlength can be shorter than
the nominal Parker spiral length, down to 1~au. The short pathlengths
are caused by the stochastic paths that are ``straightened'' from the
Parker spiral shape to radial shape. It should be noted though, that
the probability of such paths is low. Also notable is the vanishingly
small probability of paths that reach heliolongitudes larger than
$\phi\sim 45^\circ$. This is consistent with the rarity of SEP events
originating from solar eruptions farther in the Eastern
heliolongitudes, East from E45.

For SEP events on the western hemisphere (between the vertical dashed
lines labelled CM and W90 in Figure~\ref{fig:alongparker_vs_pathlen}),
the shortest pathlengths are still of the order of or shorter the
nominal Parker spiral length, 1.15~au, and only sources far behind the
western limb (left of vertical dashed line labelled W90) have
substantially longer shortest pathlengths. Thus, the onsets of even
some backside events could result in close to nominal pathlengths. The
mean pathlength, however, increases significantly for western
hemisphere and behind-the-limb sources.

As discussed in Section~\ref{sec:stat-eval-length}, the mean length of
a meandering path can be estimated using the statistical values of the
meandering path. For Parker geometry, such estimation is not as simple
as in the cartesian case, as the ``direct path'' with length $s_0$ in
Equation (\ref{eq:fulllengthvsturb}), or the statistical distribution
of steps within the meandering path, cannot be determined
unambiguously. We approach the estimation by evaluating the direct
path with a Parker spiral that joins the source longitude
$\phi_0=0^\circ$ to a longitude $\phi_r$ at a given distance $r$. Such
\REVaAdd{an undisturbed }
Parker spiral, parametrised with $a=r/\phi_{r}$, has length of
\begin{equation}
  \label{eq:parkerdirectpath}
  s_0(r, \phi_r)=\frac{r}{2\phi_r}\left[\phi_r\sqrt{1+\phi_r^2}+\ln\left(\phi_r+\sqrt{1+\phi_r^2}\right)\right].
\end{equation}
\REVaDel{With this estimation, the mean length of a path meandering
  stochastically in Parker spiral geometry,}\REVaAdd{As we saw in
  Section~\ref{sec:stat-eval-length}, in the case of cartesian
  geometry an undisturbed path experienced lengthening by a factor of
  $(1+\rho\,\delta B^2/B^2)$ due to stochastic wandering (Equation
  (\ref{eq:fulllengthvsturb}) with $\rho=Z^4/s^4\in[0, 1]$). Applying
  similar statistical lengthening to the undisturbed Parker spiral
  length, given by Equation~(\ref{eq:parkerdirectpath}), thus we can
  write the mean length of a stochastic path} from a point source at
the Sun at $\phi=0$ to $(r, \phi_r)$ \REVaDel{is given by
}\REVaAdd{as}
\begin{equation}
  \label{eq:parkerestimate}
  \left<s(r, \phi_r)\right> = s_0(r, \phi_r)\left[1+\rho
    \frac{dB^2}{B^2}\right]. 
\end{equation}
\REVaDel{where $\rho=Z^4/s^4\in[0, 1]$ (see
Section~\ref{sec:stat-eval-length}).}

Evaluating $\left<dB^2/B^2\right>=0.16$ between 1/215~au and 1~au
for our turbulence model\footnote{Note that this value differs from
  \cite{LaEa2016parkermeand} where the diffusion coefficient was
  calculated via integrating the spectrum, whereas here we use the
  unnormalised $\tilde\lambda\sim\sqrt{\lambda_c L}$ from
  \citet{Matthaeus2007}, see Appendix~\ref{sec:dfl_analytic}.}, we
show Equation~(\ref{eq:parkerestimate}) in
Figure~\ref{fig:alongparker_vs_pathlen}(c) with solid blue curve for
$\rho=0$ and dashed black curve for $\rho=1$. As can be seen, the
black curve traces well the most likely pathlengths for western
sources (left from W60, magenta symbol), whereas the eastern
pathlengths tend to be shorter, closer to the $\rho=0$ curve.

\section{Discussion}\label{sec:discussion}

In this study, we have investigated the pathlength of SEPs propagating
along interplanetary magnetic field lines that spread stochastically
across the mean magnetic field due to fieldline random walk. As we
demonstrate in Section~\ref{sec:models}, diffusion description of such
motion neglects the effect of the stochastic cross-field motion in
evaluation of distance the particle can propagate in a given time,
resulting in erroneus first-arrival time of SEPs to a given
distance. We introduced ultrascale, the scalesize of the turbulent
islands \REVaAdd{\citep{Matthaeus1999}}, as the physically-justified
characteristic scale of the fieldline meandering, and used this
concept to evaluate the pathlength of the meandering fieldline. The
resulting pathlengths are realistic, and do not exhibit the break of
causality discussed in \citet{Strauss2015}.

It is important to note that the length of the path travelled by the
SEPs is not the only problem encountered when applying timing analysis
methods such as the VDA for SEPs. The evolution of SEP intensities in
the interplanetary space is a combined effect of the length of the
meandering paths, scattering of the particles along the path
\REVaAdd{\citep{Lint04,Saiz2005}}, and propagation across the
meandering fieldlines due to diffusive escape from a path to another
\citep{LaDa2017decouple} and drifting due to the large scale curvature
and gradients of the background Parker spiral magnetic field
\citep{Dalla2013}. This is compounded with the pre-event background
intensities \citep{Laitinen2015onset}, which make it difficult to
determine when the ``first non-scattered'' particles would have
arrived.

Thus, while the shortest pathlengths in
Figure~\ref{fig:alongparker_vs_pathlen}~(c) at W60 are around 1~au, it
may be that the number of particles propagating at the low-probability
short paths are not seen above the pre-event background. Similarly,
while the mean pathlength at $\sim 225^\circ$ (W135, behind the
western limb) is around 2~au, the first observed particles may have
traversed the shorter paths with similar or only slightly lower
probability. At large heliolongitudinal distances from the
best-connected site (W60), the first-observed SEPs may have propagated
across the fieldlines due to diffusive escape \citep{LaDa2017decouple}
and drifts \citep{Dalla2013}, instead of having propagated directly
from the solar source along the meandering fieldlines. All of these
factors contribute to uncertainties in SEP timing methods such as the
VDA.

For full understanding of SEP propagation, one should thus combine the
analysis of meandering paths and the SEP transport into one
framework\REVaAdd{, to amend the often-used diffusion-convection
  approach used in many SEP and galactic cosmic ray studies
  \citep[e.g.][]{Zhang2009,Droge2010,Strauss2011, Strauss2015,Wang2015}}. In
  such a framework, the meandering pathlength should be evaluated
  using the Gaussian random walk approach introduced in this study,
  and the propagation time of simulated pseudoparticles should be
  rescaled by factor $\Delta s/\Delta z$, so that the particle with
  velocity $v$ would be able to physically take a step
  $\Delta s= v \Delta t$ in time $\Delta t$. This was partly done in
  \citet{LaEa2016parkermeand}, where particles propagated along
  stochastically meandering fieldlines, with additional spatial
  diffusion from the meandering path, but without the rescaling of the
  propagation time. In future work, we will incorporate the
  time-rescaling to the \citet{LaEa2016parkermeand} model.

\section{Conclusions}\label{sec:conclusions}

In this paper, we have discussed the problem of calculating the time
that diffusively propagating particles take to travel from their
source to the observer, noting that such evaluation cannot be provided
by the standard spatial diffusion approach. We have shown that
pathlengths derived using the \emph{SDE steplength} are very sensitive
to the selected step size, and thus not physical. We introduced the
concept of Gaussian random walk of magnetic field lines with
physically-justified step lengths as derived from the turbulence
ultrascale to provide an estimate for the distribution of
pathlengths. This approach, when applied to a Parker spiral
configuration, produces pathlengths that are consistent with
observations. We find that at 1~au for Parker spiral with solar wind
velocity of 430~km/s, the shortest pathlengths are close to the
nominal Parker spiral length, or even shorter, for a large range of
heliolongitudes, corresponding to SEP events taking place at E45 to
W90 solar longitudes when viewed from Earth. The mean pathlength
increases roughly linearly from the nominal 1.15~au for SEP events
originating at W60 to far beyond the western limb. Our method should be
used to correct for propagation time in all spatial diffusion SDE
codes, when the physical scales for the underlying random walk process
can be estimated.

\acknowledgements{ TL and SD acknowledge support from the UK Science
  and Technology Facilities Council (STFC) (grant ST/R000425/1), and
  the International Space Science Institute as part of international
  team 297. Access to the University of Central Lancashire's High
  Performance Computing Facility is gratefully acknowledged.}


\appendix

\section{Analytic expression for fieldline diffusion coefficient}\label{sec:dfl_analytic}

\REVaAdd{\citet{LaEa2016parkermeand} used a spectrum with a flat
  spectrum at scales between the largest scale in the spectrum, $L$, and the bendover
  scale $\lambda_c$, and Kolmogorov spectrum at scales smaller than $\lambda_c$.}
As discussed in \citet{Matthaeus2007}, for \REVaDel{a spectrum used in
Laitinen~et~al.~(2016)} \REVaAdd{ such a spectrum} the ultrascale is given as
$\tilde\lambda\sim\sqrt{\lambda_c L}$\REVaDel{, where $\lambda_c$ is the bendover
scale of the turbulence spectrum, and $L$ is the largest scale in the
fluctuations.} \citet{LaEa2016parkermeand} took $L\propto r$, the
radial distance from the Sun, a natural choice in a
spherically-expanding, outflowing turbulent plasma.

For the turbulence amplitude \citet{LaEa2016parkermeand} used the WKB
approximation,
\begin{equation}\label{eq:wkb}
  \delta B^2(r)=\delta B^2(r_0)\left(\frac{r_0}{r}\right)^3\left(\frac{V_{sw,0}+v_{A0}}{V_{sw,0}+\frac{r_0}{r}v_{A0}}\right)^2,
\end{equation}
where $V_{sw}$ and $v_{A}$ are the solar wind velocity and Alfv\'en
velocity, and values subscripted with 0 are those at reference
distance $r=r_0$. The Parker spiral magnetic field magnitude is given byy
\begin{equation}\label{eq:parkerb}
  B(r)=B_0\left(\frac{r_0}{r}\right)^2 \sqrt{\frac{r^2+a^2}{r_0^2+a^2}},
\end{equation}
where $a$ is the Parker spiral parameter.

Using Equations~(\ref{eq:dfl}), (\ref{eq:wkb}), (\ref{eq:parkerb}) and
$\tilde\lambda\propto\sqrt{r}$, we can write for the field-line
diffusion coefficient
\begin{equation}
  \label{eq:dflparker}
  D_{FL}(r)=D_{FL,0}\frac{r}{r_0}\frac{V_{sw,0}+v_{A0}}{V_{sw,0}+\frac{r_0}{r}v_{A0}}\sqrt{\frac{r_0^2+a^2}{r^2+a^2}}.
\end{equation}
We use as reference values at $r_0=1$~AU $v_{A0}=30$~km/s,
$V_{SW}=430$~km/s and $a=-1$. For the fieldline diffusion coefficient at 1~au, we use $D_{FL,0}=r_0^2(10^\circ)^2/$AU, consistent with
\citet{LaEa2016parkermeand}.

\end{document}